\documentclass{egpubl}
\usepackage{eg2026}
 
\JournalPaper         
\CGFccby

\usepackage[T1]{fontenc}
\usepackage{dfadobe}  

\BibtexOrBiblatex
\electronicVersion
\PrintedOrElectronic
\ifpdf \usepackage[pdftex]{graphicx} \pdfcompresslevel=9
\else \usepackage[dvips]{graphicx} \fi

\usepackage{egweblnk} 

\usepackage{hyperref}
\usepackage{amsmath}
\usepackage{amssymb}
\usepackage{graphicx}
\usepackage{comment}
\usepackage{kbordermatrix}
\usepackage{multirow,bigdelim}
\usepackage{cleveref}
\usepackage{dsfont}

\usepackage{array}
\usepackage{wrapfig}

\usepackage{csvsimple}
\usepackage{adjustbox}
\usepackage[percent]{overpic}

\usepackage{color}
\usepackage{soul}
\usepackage{breqn}
\usepackage{booktabs}
\usepackage{tabularx}
\usepackage{rotating}

\definecolor{purple}{rgb}{0.99,0.2,0.72}
\definecolor{blue}{rgb}{0, 0.2, 0.8}
\definecolor{orange}{rgb}{0.6, 0.6, 0}
\definecolor{red}{rgb}{0.8, 0.2, 0.2}
\definecolor{magenta}{rgb}{0.5, 0.0, 1.0}
\definecolor{black}{rgb}{0.0, 0.0, 0.0}
\definecolor{cyan}{rgb}{0, 0.65, 0.65}
\definecolor{olive}{rgb}{0.2, 0.6, 0.5}
\definecolor{revgreen}{rgb}{0.0, 0.5, 0.2}
\usepackage{xcolor}

\newcommand{\tpts}{{three-point }}
\newcommand{\ttpts}{{the three-point }}

\newcommand{\X}{ \textbf{X} }

\newcommand{\R}{ \mathds{R} }
\newcommand{\C}{ \mathds{C} }
\newcommand{\CC}{ \textbf{C}}

\newcommand{\Rot}{R}
\newcommand{\RR}{\textbf{R}}
\newcommand{\Pos}{P}
\newcommand{\PP}{\textbf{P}}
\newcommand{\Y}{\textbf{Y}}

\newcommand{\F}{F}
\newcommand{\FF}{\textbf{F}}

\newcommand{\head}{\text{head}}
\newcommand{\lhand}{\text{lhand}}
\newcommand{\rhand}{\text{rhand}}

\newcommand{\Loss}{\mathcal{L}}
\newcommand{\ttt}[1]{\texttt{#1}}

\newif\ifdraft
\draftfalse

\ifdraft
    \newcommand{\ssc}[1]{{\color{magenta}[\textbf{Sebastian:} \textit{#1}]}}
    \newcommand{\yyc}[1]{{\color{olive}[\textbf{Yuting:} \textit{#1}]}}
    \newcommand{\plc}[1]{{\color{cyan}[\textbf{Peizhuo:} \textit{#1}]}}
    \newcommand{\OSHc}[1]{{\color{purple}[\textbf{Olga:} \textit{#1}]}}

\else
    \newcommand{\ssc}[1]{}
    \newcommand{\yyc}[1]{}
    \newcommand{\plc}[1]{}
    \newcommand{\OSHc}[1]{}

\fi

\usepackage{pgfplots}
\pgfplotsset{compat=newest}
\usepgfplotslibrary{groupplots}
\usepgfplotslibrary{dateplot}

\usepackage[bottom]{footmisc}
\usepackage{units}
\title{Dancing Points: Synthesizing Ballroom Dancing with Three-Point Inputs}

\teaser{
\centering
    \includegraphics[width=0.9\linewidth]{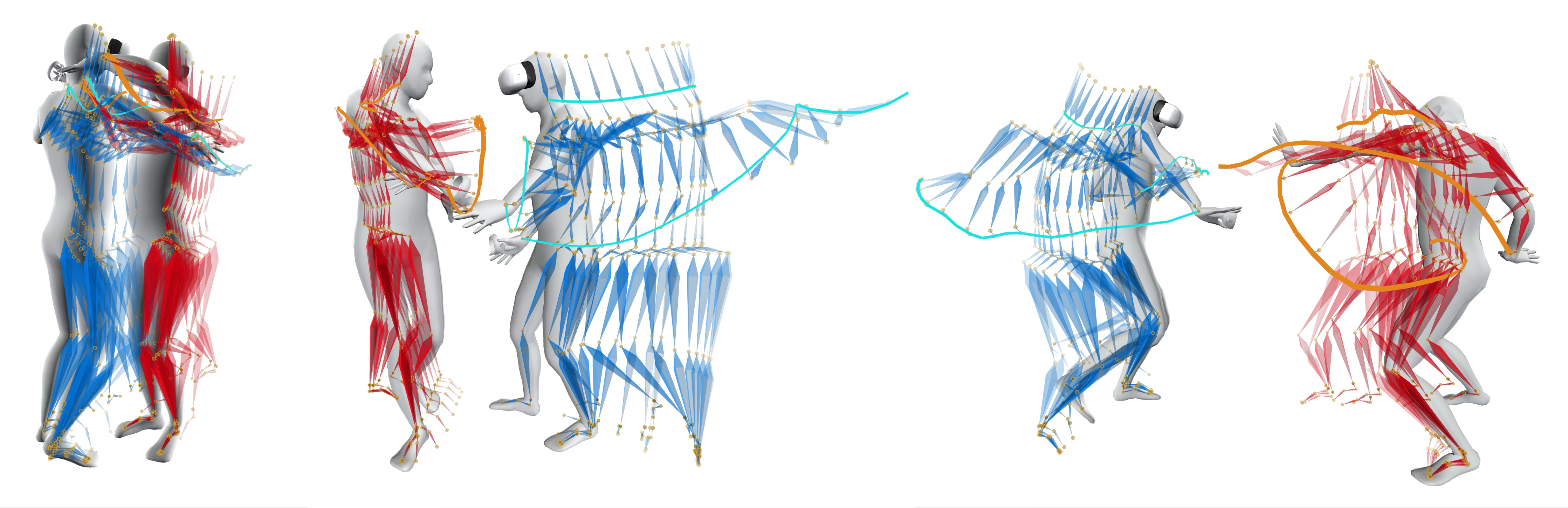}
    \caption{We synthesize full-body ballroom dancing solely from the three-point inputs captured by the VR headset on the leading dancer in real time, for various interaction scenarios, such as closed position, one-hand contact and intensive movement without body contact.}
    \label{fig:teaser}
}

\author[Li, P.\ et al.] 
{\parbox{\textwidth}{\centering Peizhuo Li$^{1}$\orcid{0000-0001-9309-9967}, Sebastian Starke$^{2}$\orcid{0000-0002-4519-4326}, Yuting Ye$^{2}$\orcid{0000-0003-2643-7457}, Olga Sorkine-Hornung$^{1}$\orcid{0000-0002-8089-3974}
        }
        \\
{\parbox{\textwidth}{\centering $^1$ETH Zurich, Switzerland $\quad \quad \quad$
         $^2$Meta Reality Labs
       }
}
}

\begin{document}

\maketitle

\begin{abstract}
  Ballroom dancing is a structured yet expressive motion category. Its highly diverse movement and complex interactions between leader and follower dancers make the understanding and synthesis challenging.
We demonstrate that the three-point trajectory available from a virtual reality (VR) device can effectively serve as a dancer's motion descriptor, simplifying the modeling and synthesis of interplay between dancers' full-body motions down to sparse trajectories. 
Building on this observation, our central design decomposes dual-character synthesis into a non-autoregressive mapping network that predicts the future three-point trajectories of both dancers from the leader's input, followed by two independent full-body tracking problems. Disentangling the dual-character interaction from the full-body autoregressive rollout is what enables stable and responsive real-time generation on a small dataset, and avoids the overfitting that arises when conditioning one autoregressive model on another. The same decomposition further reduces the per-step ambiguity of three-point tracking, so the tracker can be simplified to a lightweight deterministic model despite the task being typically considered under-constrained and addressed with generative models. We further show that the design generalizes beyond small, structured ballroom dancing datasets and performs robustly on larger, more diverse datasets such as LaFAN~\cite{harvey2020robust}.
Our method provides a computationally- and data-efficient solution, opening new possibilities for immersive paired dancing applications.


\printccsdesc   
\end{abstract}

\section{Introduction}

Ballroom dancing is one of the more challenging motion categories, characterized by the intricate interplay between the leader and the follower.
Dancers strive to achieve harmonious coordination, creating movements that are both synchronized and dynamic. The leader initiates and shapes the movement, setting the phrasing and rhythm, while the follower responds seamlessly, mirroring the leader's intent to maintain continuity and flow.
Modeling and synthesis of such interaction with dynamic motion and maintaining a seamless flow is challenging, particularly when both dancers need to be generated in real time based on online input from a VR headset.

The structure and dynamic movements in ballroom dancing present both challenges and, interestingly, advantages when combining with three-point tracking tasks. Predicting full-body motion in real time with three-point inputs is known to be difficult because of the sparse yet strict constraints (e.g., the exact positions of the head and hands) and is inherently ambiguous due to the absence of lower-body information.
The former generally requires a carefully designed autoregressive model that responds swiftly to the changing input signal with smooth transition.
At the same time, the ambiguity is reduced, since in ballroom dancing the movements of the hands and upper body are more informative and aligned with the lower body. This allows us to use the three-point trajectories as an explicit, compact yet informative descriptor for a dancer's movement.
Since three-point trajectories are of lower dimension, modeling the interaction between them is significantly easier comparing to handling full-body motion. In fact, when the dancers are executing the \emph{passive follow} paradigm -- the follower's act of moving only in direct response to a clear lead without anticipating or initiating motion-- it is possible to synthesize plausible full-body motions for both dancers, by first modeling the three-point trajectories and solving the two subsequent three-point tracking problems independently, bypassing the need to condition the follower's motion synthesis on the predicted full-body leader motion. Being able to avoid conditioning on the output of another autoregressive model is critical for stable and responsive rollouts.

Based on this observation, our central design \emph{decomposes} dual-character synthesis into a non-autoregressive mapping network that predicts the future three-point trajectories of both dancers from a short window of past three-point inputs, followed by two independent three-point tracking problems that recover each dancer's full-body motion in real time.

While a passive-follow assumption may appear restrictive, it realistically reflects the inherent structure of many partnered dance styles and aligns well with user experience in VR. In a broad range of ballroom dances (e.g., Waltz, Quickstep, Foxtrot, classical Tango), the roles of leader and follower are clearly defined and not intended to switch. From the user's perspective, they directly control the leader's motion, and an effective model should therefore generate follower motion that is responsive rather than contradictory. Combining these considerations, our formulation emerges as a practical application suited to the constraints of three-point input in VR. Similar interaction concepts have also been informally explored in the VR gaming community~\cite{redditDanceVR}.

While unconditional or text-conditioned motion synthesis has seen significant success with the advent of diffusion models and large datasets, three-point tracking remains a persistent challenge.
Techniques such as codebook matching~\cite{starke2024categorical} attempt to address the inherent ambiguity by identifying nearest neighbors in a discrete latent space. However, this approach suffers when the training data is limited, especially for complex motion like ballroom dancing.
Our decomposition addresses this differently: by disentangling the dual-character interaction from the full-body autoregressive rollout, the mapping network supplies each tracker with a predicted future \tpts trajectory, and conditioning the tracker on that trajectory together with the current pose makes the one-step advance close to unimodal, so the tracker itself can be a deterministic model rather than a generative one. The decomposition is broadly effective: it produces robust and plausible results both on our small, structured ballroom dancing dataset and on the much larger and more diverse LaFAN~\cite{harvey2020robust} locomotion dataset. While the predicted motion may not be identical to the ground truth due to the inherently limited input information, it remains coherent and realistic.

Another benefit of predicting the pair of ballroom dancers via three-point trajectories is generalization. The low-dimensional nature of the three-point trajectory alleviates the need for a large dataset and a learned latent representation, reducing the risk of overfitting and ensuring the robustness of our model when big data is not available.

We capture a small-size ballroom dancing dataset and showcase the ability of our framework on it. By combining the three-point future predicting network with full-body motion tracking network, we predict the full-body motions of both the leader and follower dancers in real time, solely based on the leader's three-point input. The system demonstrates high responsiveness to changing input signals, enabling harmonious and dynamic interactions between the two dancers.

\section{Related Work}
Deep learning has been demonstrated as a powerful tool for a wide range of applications in kinematics-based animation \cite{holden2017phase, zhang2018mode, starke2020lmp, henter2020moglow, tevet2023human, jiang2023motiongpt, 2023PhaseBetweener} and physically-based animation\cite{2018-TOG-deepMimic, 2020MotionVAE, won2022cvae}. As our work mainly focuses on VR-based dance interaction synthesis, the following sections primarily mention works most related to our research.

\begin{figure*}
    \centering
    \includegraphics[width=\linewidth]{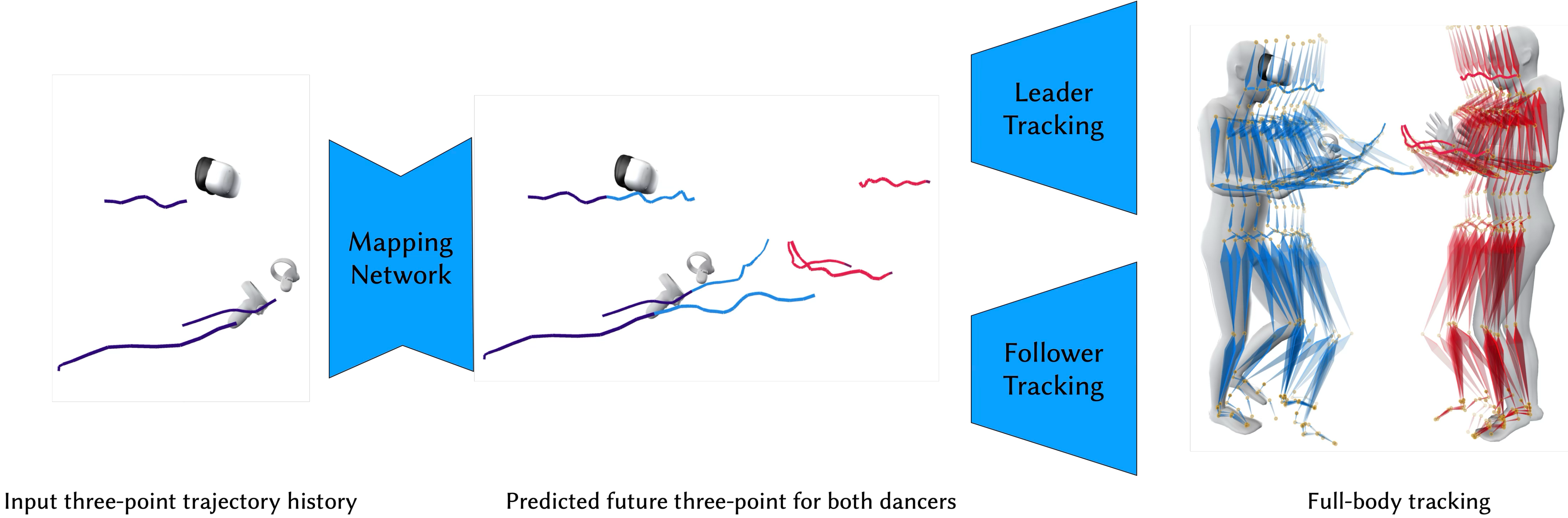}
    \caption{Our framework begins with the \tpts input from the past 0.5 seconds of the leading dancer. The mapping network maps this \textcolor[rgb]{0.180, 0.027, 0.475}{input} to the future \tpts positions for both the \textcolor[rgb]{0.118, 0.533, 0.898}{leader} and the \textcolor[rgb]{0.898, 0.118, 0.306}{follower}. Using these predicted \tpts positions, along with the pose predicted from the previous frame, the tracking network autoregressively predicts the future root trajectories and the future full-body motions.}
    \label{fig:arch}
\end{figure*}

\subsection{Three-point tracking}

With the growing interest in virtual reality applications, the three-point tracking problem has become increasingly prominent. Its goal is to estimate the full-body movements of an avatar from a set of sparse sensor inputs. Early works \cite{jiang2022avatarposer} using a transformer-based model can suffer from smoothing artifacts due to the deterministic nature of the method.

    To address the ambiguity issue, additional sensors on the lower body are used to provide more information~\cite{yang2021lobstr, zeng2022pe, ponton2023sparseposer}.
    Another approach is to use a diffusion model to handle ambiguity~\cite{du2023avatars, castillo2023bodiffusion, zhang2024rohm}. However, these methods are only suitable for offline tracking due to the requirement of future control signal input and high computational cost.

Alternatively, Ponton et al. \cite{ponton2022mmvr} apply motion matching to achieve more robust transitions, but this method may struggle in terms of scalability for larger datasets and has high tracking error. Winkler et al. \cite{lee2023questenvsim} utilize reinforcement learning to generate physically-based movements for tracking diverse user movements and interactions, but require the motion to be synthesized with temporal delay. To overcome such situations, Starke et al. \cite{starke2024categorical} utilize a discrete probabilistic approach by projecting the motions against valid samples in a learned categorical codebook. However, what is shared across those methods is that they remain limited to single-character motions and lack exploring the spatial or temporal interaction between multiple avatars. Our work specifically focuses on this open problem in VR motion tracking.

\subsection{Multi-character interaction}

In Starke et al. \cite{starke2021neural}, the problem of close-character interactions is addressed in the context of fighting and clinging animations. A neural network is used to transform combinations of kinematic reference trajectories for different body parts into novel movements, and a contact matrix is predicted to resolve contact pairs between interacting body limbs. In \cite{zhang2023interaction}, a model is trained to identify physically plausible interactions between characters and when interacting with objects. Other works address the problem of modeling crowd-based animations for contact response behaviors \cite{shum2008patches} or realistic navigation during locomotion \cite{greilcrowds2023}. Although these works share many similarities in their challenges to our research, our ballroom dancing motions also require synchronized and dynamic responses.
Recently, diffusion models have been employed to synthesize reactive motions conditioned on the motion of a leading character \cite{chopin2023interaction,ghosh2024remos,xu2024regennet}. Similarly, synthesizing two-person interaction motions directly from text prompts are also made possible using diffusion \cite{liang2024intergen, javed2024intermask, tan2025think}. Siyao et al. \cite{siyao2024duolando} propose a hybrid approach that combines a GPT-style architecture with reinforcement learning to synthesize the follower's motion in partner dancing scenarios. While effective, these methods assume that conditioning information—such as the leader's motion or a text description—is available in advance, and they typically involve substantial computational overhead due to the nature of diffusion-based generation. In contrast, our approach is designed for real-time applications, where input signals arrive online and demand low-latency, responsive motion synthesis.

Recently, Ready-to-react \cite{cen2025ready} propose an online model focuses on boxing motion. Ji et al.\  \cite{ji2025towards} build a real-time framework that responses to various type of input signals combined with a text prompt, and synthesize motion with physically-based tracking method. However, those works does not address the accurate close-body cooperation required in partner dancing.

\subsection{Dancing motion}

Various works explore generating realistic dance motions. A key challenge with dance motions is that they can greatly differ in timing and synchronization in how different body parts move together, compared to standard locomotion patterns. In \cite{kang2021choreomaster, starke2022deepphase, alexanderson2023lda}, dance behaviors are generated given the music as input, where animations should closely match the rhythm using motion graphs \cite{kang2021choreomaster}, learned phase variables \cite{starke2022deepphase} or diffusion-based approaches \cite{alexanderson2023lda}. Wu et al.\ \cite{wu2025transformer} propose a transformer-based approach for teaching partner dancing with an emphasis on Laban movement analysis. In \cite{tseng2022edge}, a physics-based model using a cVAE architecture is trained to generate diverse dance motions. DuetGen~\cite{ghosh2025duetgen} generates dancing motion directly from music with a quantized two-person motion representation. Different from these works, our method tackles the problem of estimating not only the motion of a following dancer given the motion of the leading dancer, but also the motion of that leading dancer itself from sparse sensor inputs.

\section{Overview}

Our goal is to generate full-body ballroom dancing motions for both the leader and the follower using the VR tracker signals from the leader's head and two hands, referred to as the three-point input, in real-time at 30 Hz on a nominal laptop.

Our motion synthesis system consists of two key components: tracking and mapping, as described in \Cref{fig:arch}. The tracking network, described in \Cref{sec:tracking-net}, predicts future full-body motion by leveraging the \emph{current pose} and \emph{future trajectory} of \ttpts in an autoregressive manner.
The mapping network, described in \Cref{sec:mapping-net}, predicts the \emph{future \tpts trajectory} of \ttpts for both the leader and the follower from the leader's \emph{past \tpts  trajectory}. Then these are fed into the tracking network to generate the corresponding future full-body motions.

\section{Method}

In this section, we describe our motion representation, the input feature and output feature design for each component, the autoregressive process and the design of our neural networks. A central theme of our design is to keep each component as simple as possible. Decomposing dual-character synthesis into a non-autoregressive mapping network that predicts both dancers' future \tpts trajectories and an independent tracking network that converts a trajectory into full-body motion, together with conditioning the tracking network on the \emph{future} trajectory, allows us to simplify the tracking network into a lightweight deterministic MLP -- a simplification of the categorical codebook formulation of CBM~\cite{starke2024categorical}.

\subsection{Motion representation}

Each frame of our motion is represented by two parts, a root trajectory component on the floor plane $\F = (t, o) \in \R^4$ in the global coordinate system, consisting of the translation $t \in \R^2$ and the orientation $o \in \C$ represented as a complex number by projecting the head position and orientation to the floor, and a pose component consisting of all the $J$ joint positions $\Pos \in \R^{J \times 3}$ and rotations $\Rot \in \R^{J \times 9}$, represented by 3D rotation matrices, in the corresponding root coordinate system relative to $\F$. We also compute the foot contact labels $C \in \{0, 1\}^4$ for the heels and toes.
A motion sequence $\{ \FF, \RR, \PP, \CC \}$ consists of root trajectory $\FF = \{\F^t\}_{t=1}^T$, joint rotations $\RR = \{ \Rot^t \}_{t=1}^T$, joint positions $\PP = \{ \Pos^t \}_{t=1}^T$ and foot contact labels $\CC = \{ C_t\}_{t=1}^T$.

We slightly abuse notation by not distinguishing the aforementioned representations or their corresponding affine transformations or coordinates. For example, $\F\Pos$ indicates the affine transform of $F$ being applied to the affine coordinates of positions $\Pos$, meaning the transform from local positions $P$ in the coordinate system defined by $F$ to the global coordinate system. Similar rules apply to the position sequence $\PP$, rotations (sequence) $\Rot$ ($\RR$) and root trajectory (sequence) $\F$ ($\FF$).

Note that we do not explicitly model the hierarchical structure of the skeleton, as this representation helps mitigate the typical error accumulation along the kinematic chain. Moreover, it is better suited for our application, since the \tpts input can be represented in the same manner, using only a subset of the complete set of joints, ensuring consistency and simplicity in the representation.

For the raw three-point input from the VR trackers, we assume it consists of orientations and positions of the head and two hands. We then calculate the root trajectory $\F$ as mentioned above, and we denote the tracker positions by $\bar\Pos = \{ \Pos_\head, \Pos_\lhand, \Pos_\rhand\} \in \R^{3 \times 3}$ and tracker rotations by $\bar\Rot = \{\Rot_\head, \Rot_\lhand, \Rot_\rhand\} \in \R^{3 \times 9}$, both relative to the root trajectory $F$.

\subsection{Tracking network}
\label{sec:tracking-net}

\begin{figure}
    \centering
    \includegraphics[width=\linewidth]{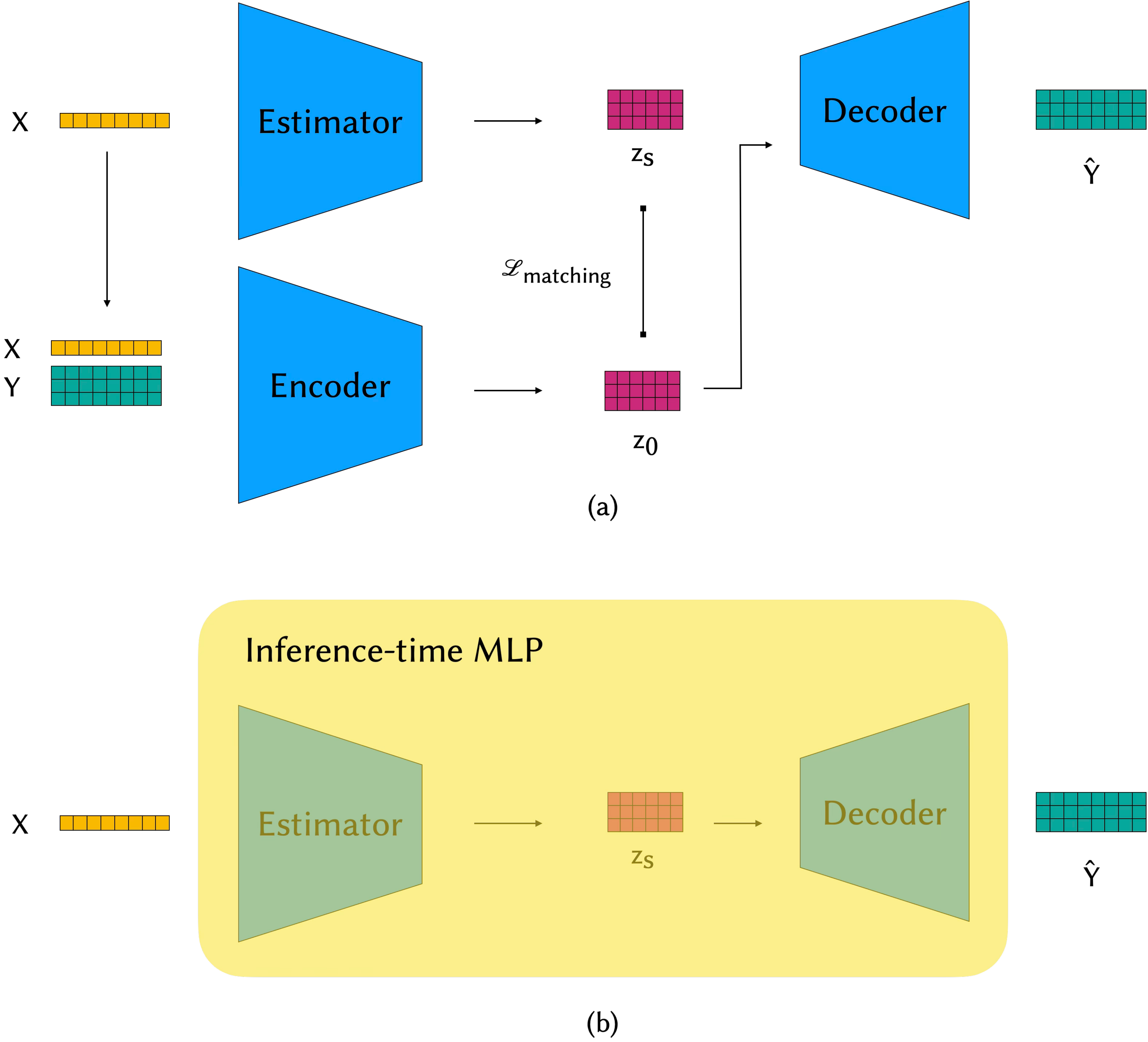}
    \caption{Overview of the tracking network. (a) During training, we use an auto-encoder to learn the latent space, and an estimator to predict the latent code $z_s$ that corresponds to the input $\X$. (b) During inference, the encoder is discarded, reducing the tracking network to an MLP.}
    \label{fig:arch-tracking}
\end{figure}

Our tracking network is a light-weight feedforward network that takes the future \tpts trajectories and the current pose as input and predicts the full-body motion of the corresponding frames.
We adopt a simplified version of Codebook Matching~\cite{starke2024categorical} that replaces the discrete codebook with a continuous latent space, trained using an estimator-encoder-decoder triplet. At inference time, the encoder is discarded and the network reduces to a deterministic MLP, as illustrated in \Cref{fig:arch-tracking}.

Given the raw signal of the future 1 second from \tpts input with root trajectories $\FF$ and three-point joint positions $\bar\PP$, we transform the root trajectories relative to the root from the last prediction $\F_0$ by $\F_0^{-1}\FF$ and concatenate $\bar\PP$ and $\F_0^{-1}\FF$, together with the current pose $\Pos_0, \Rot_0$ and foot contact labels $C_0$ as the input of the tracking network, denoted by $\X = \{ \Pos_0, \Rot_0, C_0, \F_0^{-1}\FF, \bar\PP \}$. To prevent overfitting, we do not use the rotations from the input signal.

The tracking network predicts the future root trajectories $\FF$, joint rotations $\RR$, positions $\PP$ and foot contact labels, denoted by $\Y = \{\FF, \RR ,\PP, \CC\}$ for the next 1 second. The predicted root trajectory is relative to the current root $\F_0$. This prediction is used to advance the character by one frame, and the process is repeated at every frame. We follow the same process as in CBM~\cite{starke2024categorical} to recover the final motion from the network output. For more details, please refer to Appendix \ref{appendix:visualization}.

We train the tracking network with the following loss:
\begin{align}
     & \Loss_{\text{rec}} = \mathbb{E}_{(\X, \Y) \sim \textbf{D}}\|\mathcal{D}_t(\mathcal{E}_t(\X)) - \Y\|_2,      \\
    \label{eq:loss-rec}
     & \Loss_{\text{matching}} = \mathbb{E}_{(\X, \Y) \sim \textbf{D}}\|\mathcal{E}_t(\X) - \mathcal{S}_t(\X)\|_2,
    \\
     & \Loss_{\text{tracking}} = \Loss_{\text{rec}} + \Loss_{\text{matching}}
\end{align}
where $\mathcal{D}_t(\cdot), \mathcal{E}_t(\cdot)$ and $\mathcal{S}_t(\cdot)$ are decoder, encoder and estimator respectively, $\textbf{D}$ is the corresponding dataset and $(\X, \Y)$ are sampled from 1 second continuous motion.

Although the overall mapping from a \tpts trajectory to full-body motion is inherently ambiguous, we show in \Cref{sec:exp-tracking} that the one-step advance of our autoregressive process given the current pose and the future \tpts trajectory can be effectively handled by a deterministic model.

Per-frame accuracy alone, however, does not guarantee stable long-term generation, as small errors can compound across the autoregressive rollout. We address this through the following designs: supervising the prediction of a full second of future motion with true future \tpts trajectories input forces the network to track the real future \tpts trajectory rather than remembering it from the past. By transforming the input future trajectory into the coordinate system of the last prediction, any root-trajectory error accumulated over the rollout manifests as a drift of the future trajectory, which in turn encourages the network to realign with the control signal. Feeding only a single frame of past prediction back into the network prevents it from reproducing future motion by memorizing past motion, the dominant source of long-term instability in autoregressive models. Discarding the rotations of the three-point input further guards against overfitting to the input signal.

\subsection{Mapping network}
\label{sec:mapping-net}

As shown in \Cref{sec:exp-tracking}, the tracking results demonstrate that the \tpts can effectively serve as a motion descriptor. To ensure stable progression in the autoregressive model, we use a separate network to reliably predict the future signals required by the tracking network, as illustrated in \Cref{fig:arch}.

Our mapping network is designed with two purposes: predicting the leader's future \tpts input to eliminate latency caused by the follower's future trajectory requirement and predicting the follower's future \tpts input to enable close and responsive interactions. Both predictions are derived from the leader's past \tpts input. We choose a non-autoregressive design for this network to ensure that its outputs remain sharp and progress steadily, allowing the tracking network to operate smoothly using the mapping network's outputs.

Formally, given the past 0.5 seconds of the past \tpts input of joint $\X = \{\FF^{\text{past}}, \bar\PP^{\text{past}}\}$, transformed into the coordinate system defined by the first frame $\F_0$, the mapping network predicts the future 1 seconds \tpts positions and root trajectory for both the leader and the follower, $\Y = \{\FF^{\text{future}}, \bar\PP^{\text{future}}\}$. This network can be trained with the following loss function:
\begin{equation}
    \Loss_{\text{mapping}} = \mathbb{E}_{(\X, \Y) \sim \textbf{D}} \| \mathcal{F}_m(\X) - \Y \|_2,
    \label{loss:mapping}
\end{equation}
where $\mathcal{F}_m(\cdot)$ is the mapping network, $\textbf{D}$ is the corresponding dataset and $(\X, \Y)$ are sampled from 1.5 seconds continuous motion.

Similarly, we find that a simple MLP is sufficient to make the prediction for the leader's future three-point trajectories (and future trajectories for LaFAN) regardless the dancing style, and for follower's future three-point trajectories when using passive following paradigm. For more evaluations, please refer to \Cref{sec:exp-mapping}.

\section{Experiments}

\subsection{Dataset and training}

 \begin{table}
     \centering
     \setlength{\tabcolsep}{4.5pt}
     \caption{Dataset details.}
     \small
     \begin{tabular}{lcccccc}
         \toprule
         Style & Balboa & Chacha & Foxtrot & Freestyle & Hustle & V-Waltz \\
         \midrule
         \small \# frames & 125,115 & 117,142 & 136,542 & 83,241 & 119,040 & 140,975 \\
         \bottomrule
     \end{tabular}
     \label{tab:dataset}
 \end{table}

We collect a dataset of ballroom dancing motions performed by professional dancers, including Balboa, Chacha, Foxtrot, Freestyle, Hustle and Viennese Waltz, with each style containing approximately from 10 to 20 minutes of motion capture data for both the leader and follower captured as a few long clips of roughly equal length as detailed in \Cref{tab:dataset}. We collect the dataset at \unit[120]{Hz}, and downsample it to \unit[30]{Hz} during training and evaluation. The dataset is captured with optical marker and solved into a skeleton with 34 joints.

    \paragraph*{Train-test split.}
    Dancing motions include repetitive patterns, so simply splitting the dataset by frames (e.g., odd/even frames) would lead to data leakage. Instead, we divide the dataset into training and test splits in an 8:2 ratio using a deterministic temporal split: for each motion style, the first 80\% is used for training and the last 20\% for testing.
    To ensure this strategy provides a robust evaluation, we analyze the feature space overlap between the splits. We use the translation- and rotation-invariant motion representation $\Y$ defined in \Cref{sec:tracking-net}, namely, 1-second motion transformed into the coordinate of the first frame.
    We compute the distribution of distances from each test sample to its nearest neighbor (NN) in the training set (Test-Train NN) within the same dancing style, measured by the L2 norm of the motion representation $\Y$.
    The mean Test-Train NN distance is 27.43, closely matching the mean cross-clip Train-Train NN distance of 26.21, which is computed between segments from different clips of the same style. The within-clip Train-Train distance is only 1.24. The Test-Train distance is as large as that between different clips and significantly larger than the within-clip distance. This shows that our 8:2 temporal split achieves a level of novelty comparable to a cross-clip split, while retaining substantially more training data.
    These results demonstrate that our test set contains distinct motion patterns not present in the training data.
    We provide additional visual comparisons between the splits in the accompanying video.

In addition to the collected dataset, we also evaluate our method for ballroom dancing synthesis on DD100~\cite{siyao2024duolando} dataset, and three-point tracking on LaFAN~\cite{harvey2020robust}. All datasets are divided into training and test splits in an 8:2 ratio. All qualitative and quantitative results shown in the paper and the accompanying video are generated on the test set. We train our neural networks on an RTX 4090 GPU by Adam optimizer with learning rate $5 \times 10^{-5}$ plus a cosine learning rate scheduler. During training, we use the positions and global rotations of the left wrist, the right wrist and the head joint as the VR headset input. For the details of the neural network architectures, please refer to Appendix \ref{appendix:arch}.

\subsection{Three-point tracking}
\label{sec:exp-tracking}

\begin{table}
    \centering
    \caption{Quantitative evaluations for \tpts tracking on the dancing dataset. Errors are in centimeters, foot contact is in F1 score.}
    \small
    \begin{tabular}{lcccc}
        \toprule
                        & Tracking err. $\downarrow$ & Full-body err. $\downarrow$ & Foot contact $\uparrow$\\
        \midrule
        AGRoL (offline) & 6.32          & 7.63           & 0.88         \\
        CBM (original)  & 16.1          & 18.7           & 0.70         \\
        CBM             & 14.2          & 16.3           & 0.71         \\
        Ours            & 5.68          & 8.32           & 0.90         \\
        \bottomrule
    \end{tabular}
    \label{tab:tracking}
\end{table}

\begin{figure}[h]
    \centering
    \includegraphics[width=\linewidth]{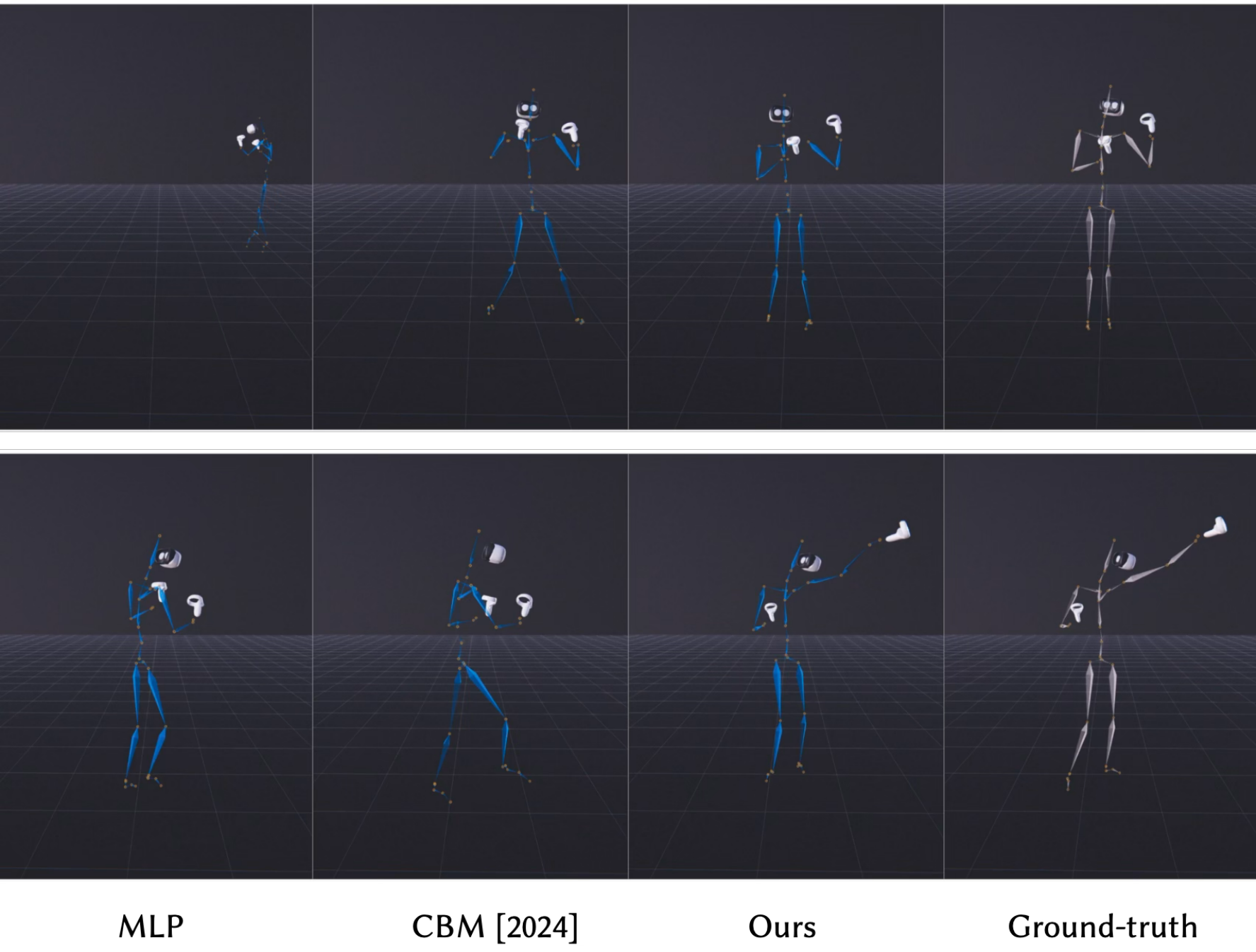}
    \caption{Comparison of three-point tracking. Each row corresponds to a frame. MLP is unstable in the autoregressive rollout. Codebook matching (CBM)~\shortcite{starke2024categorical} has difficulty fully capturing the motion. Our model captures details and faithfully reflects the movement of the three-point input.}
    \label{fig:tracking}
\end{figure}

We demonstrate that our model can synthesize leader's full-body motion from three-point inputs, producing plausible results that resemble the ground-truth pose as shown in \Cref{fig:tracking}. We compare our method against codebook matching (CBM)~\cite{starke2024categorical}, the current state-of-the-art for real-time three-point tracking, and Avatar Grows Legs (AGRoL)~\cite{du2023avatars}, an offline three-point tracking method. CBM has difficulty fully capturing the motion. We suspect this reflects a combination of factors: its discrete latent space restricts the granularity of representable motions, and training such a generative model with limited data is more demanding than fitting our deterministic mapping. This is reflected in \Cref{tab:tracking}, which reports the tracking and full-body error and the foot contact F1 score. The tracking error is the average position error of the three \emph{tracked points}, the head and the two wrists, that are directly provided by the three-point input, whereas the full-body error is the average position error over \emph{all joints}. Both are reported in centimeters. We evaluate two CBM variants: the original implementation and a ``mixed'' version, where our network architecture is replaced with CBM while retaining our input and output features. A vanilla MLP baseline is omitted from comparison since it fails to produce stable rollouts. Without specification, CBM denotes the mixed version, which consistently outperforms the original implementation. Although AGRoL slightly outperforms our method in full-body error, it is an offline method requiring an entire sequence of three-point inputs before making predictions, and is impractical for real-time applications.

\begin{figure}[t]
    \centering
    \includegraphics[width=\linewidth]{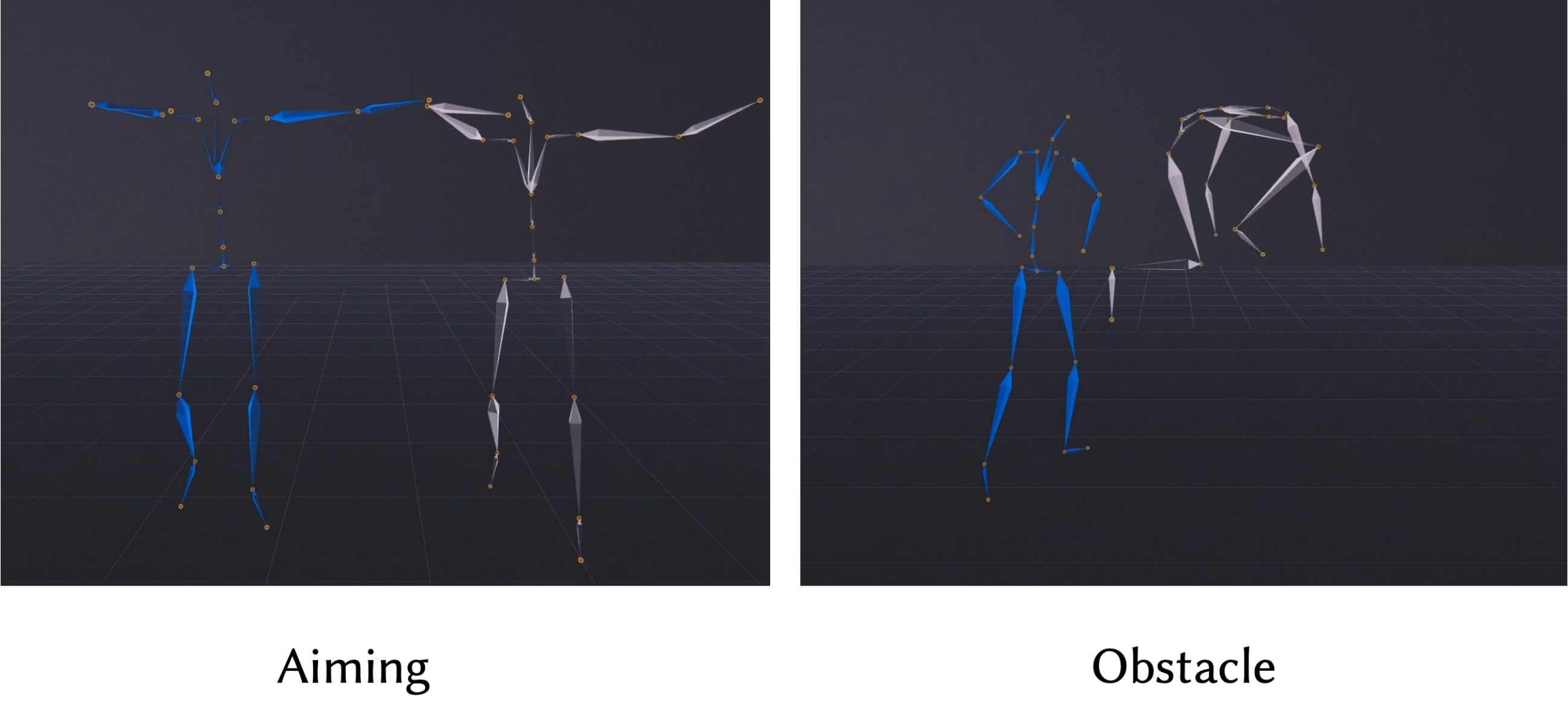}
    \caption{Three-point tracking on LaFAN~\shortcite{harvey2020robust} dataset. Our prediction (blue) follows the three-point input, though it may not match the ground-truth (white) exactly due to dataset diversity.}
    
    \label{fig:lafan}
\end{figure}

We also evaluate our model on the LaFAN~\cite{harvey2020robust} dataset. Despite the dataset's diversity, our model robustly predicts full-body motion rollouts. As shown in \Cref{fig:lafan}, the predicted motions are typically not identical to the ground truth due to the limited information available from three-point input. Nevertheless, they remain plausible and accurately follow the input trajectories. \begin{table}
    \centering
    \caption{Quantitative evaluations for tracking on LaFAN dataset. Errors are in centimeters, foot contact is in F1 score.}
    \small
    \begin{tabular}{lccc}
        \toprule
                        & Track. err. $\downarrow$ & Root drift $\downarrow$ & Foot contact $\uparrow$    \\
        \midrule
        AGRoL (offline) & 0.36          & 3.21       & 0.88         \\
        CBM (original)  & 0.38          & 5.32       & 0.80         \\
        CBM             & 0.29          & 3.96       & 0.80         \\
        Ours            & 0.31          & 3.01       & 0.81         \\
        \bottomrule
    \end{tabular}
    \label{tab:lafan}
\end{table}

\begin{figure}
    \centering
    \includegraphics[width=\linewidth]{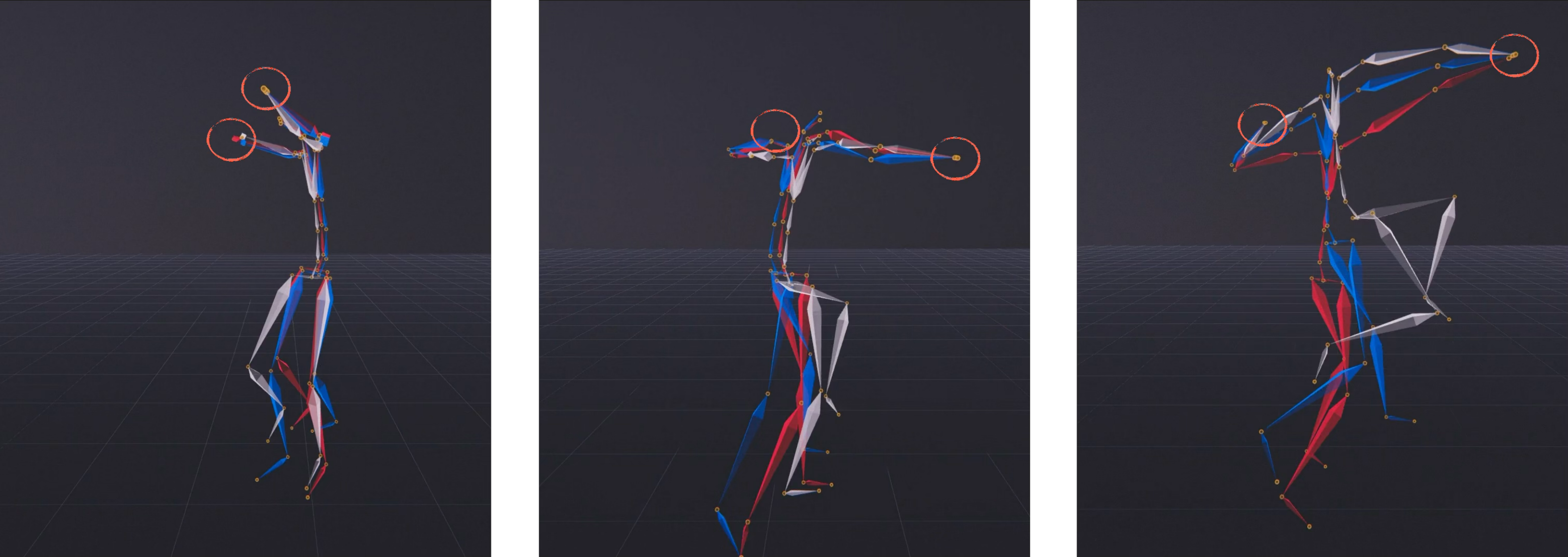}
    \caption{Comparison of {\color{blue}{our method (blue)}} and {\color{red}{codebook matching (red)}} on the LaFAN dataset, overlayed with the ground truth (white). The hand tracking errors are small despite the predicted motion not being identical to the ground truth.}
    \label{fig:lafan-overlay}
\end{figure}

We demonstrate a quantitative comparison in \Cref{tab:lafan}. For fairness, we use the same post-processing to lock the three-point signal. Note that foot-contact F1 is not strictly monotonic, as plausible motions may differ from the ground-truth labels.
For both our method and CBM, as shown in \Cref{fig:lafan-overlay}, the average tracking error is less than \unit[1]{cm}, making the three tracked points closely aligned with the ground truth most of the time, while the full body may still deviate visibly because the problem is under-constrained. Similarly, the root drift remains minimal and is negligible in practice.
Following the evaluation convention of CBM~\cite{starke2024categorical}, before computing the tracking error we apply an inverse-kinematics post-processing step that optimizes the joint rotations so that the three tracked points better match the input signal, making our numbers directly comparable to those reported by CBM.
For a comprehensive qualitative comparison, please refer to the accompanying video.

\subsection{Follower synthesis}
\label{sec:exp-mapping}

\begin{table}
    \centering
    \caption{Quantitative evaluations for the follower synthesis. Errors are in centimeters, foot contact is in F1 score.}
    \small
    \begin{tabular}{lcccc}
        \toprule
              & Full-body err. $\downarrow$ & Foot contact  $\uparrow$ & Hand err. $\downarrow$ & Root corr. $\uparrow$ \\
        \midrule
        GT    & 0              & 1            & 0         & 0.82       \\
        CAMDM & 32.7           & 0.65         & 63.42     & 0.15       \\
        Ours  & 9.21           & 0.90         & 10.54     & 0.80       \\
        \bottomrule
    \end{tabular}
    \label{tab:mapping}
\end{table}

\begin{figure}[t]
    \centering
    \includegraphics[width=0.9\linewidth]{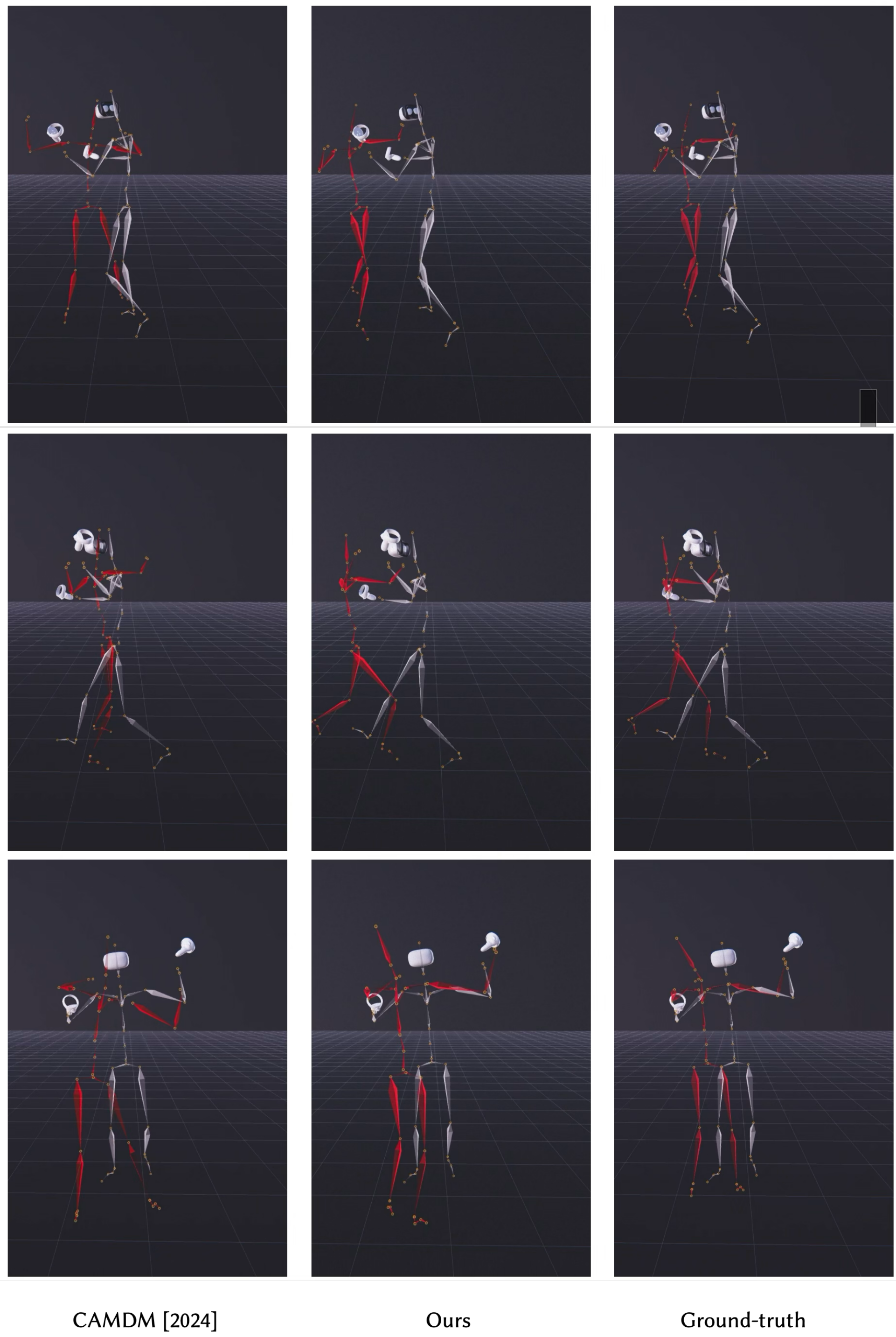}
    \caption{Comparison for follower dancer prediction. The predicted follower is in red. CAMDM's prediction {\protect \cite{chen2024taming}}  predominantly relies on the prediction of previous frames, which leads to error accumulation and overfitting. Our model is responsive to the input signal and produces motion that naturally interacts with the leader.}
    \label{fig:follower}
\end{figure}

We demonstrate that our model can synthesize follower's motion that interacts naturally with the leader dancer. We compare our method with the conditional autoregressive motion diffusion model (CAMDM)~\cite{chen2024taming}, which is the state-of-the-art in real-time motion control equipped with a diffusion model.

    We evaluate the follower synthesis using several metrics in \Cref{tab:mapping}.
    \textit{Full-body error} and \textit{hand error} measure the average joint position error and hand-to-hand distance during contact, respectively, both in centimeters.
    \textit{Foot contact} is evaluated using the F1 score.
    \textit{Root correlation} measures the circular correlation coefficient between the root headings of the leader and the follower.

It can be seen in \Cref{fig:follower} and \Cref{tab:mapping} that CAMDM is less responsive to the input signals and tends to synthesize motion predominantly based on the past motions due to its design.
Please refer to the accompanying video for further qualitative results.

Thanks to the simple network architecture, our model is computationally efficient. When evaluated with Unity's ONNX runtime library without any dedicated performance tuning, our model runs at \unit[7.8]{ms} per frame per character on Apple M1 CPU, whereas CAMDM~\cite{chen2024taming} with 4 diffusion steps requires \unit[28]{ms}, and needs a RTX 3090 GPU for real-time performance. Since our model responds immediately to incoming signals without buffering, the end-to-end latency is effectively limited to the frame time. In a 5-minute stress test streaming from a PC (AMD Ryzen 9 5900HX) to a Quest 3 headset, we achieve 30 fps for 95.3\% of the time, with an average VR-handling overhead of \unit[8.6]{ms} and per-character prediction of \unit[7.1]{ms}.

\begin{figure}
    \centering
    \includegraphics[width=\linewidth]{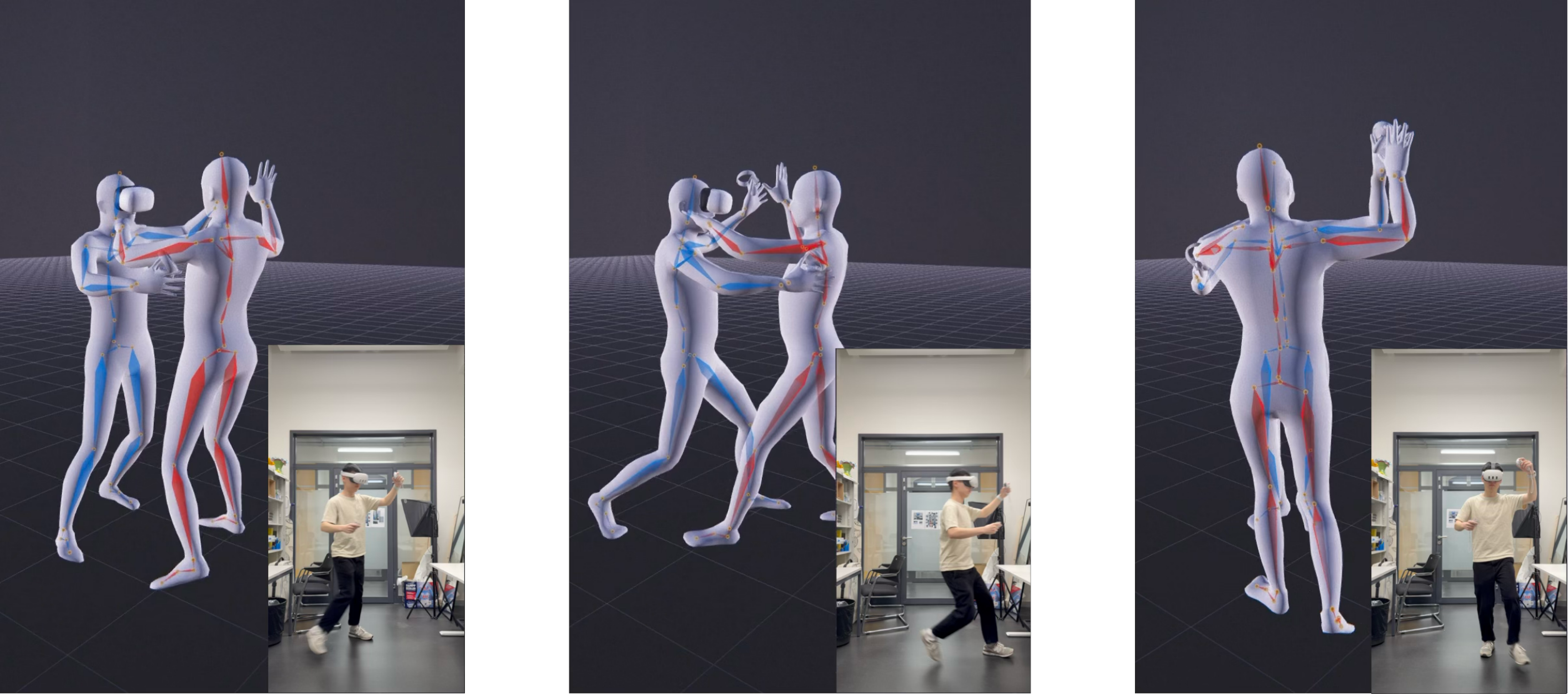}
    \caption{Example of an amateur user wearing a Quest 3 headset, performing a sequence of ballroom dancing, and our system successfully tracking the {\color{blue}{leader's motion (blue)}} and predicting the {\color{red}{follower's motion (red)}}.}
    \label{fig:embodied}
\end{figure}

    We showcase an amateur user wearing a Quest 3 headset, performing a sequence of ballroom dancing, and our system successfully tracks the leader's motion and predicts the follower's motion, as shown in \Cref{fig:embodied}.

\subsection{Ablation study}
\label{sec:ablation}

We study variants on the mapping network and demonstrate the importance of using the three-points for learning the interaction between the two dancers.

    \paragraph*{Predicting horizon} The future 1-second tracking and past 0.5-second mapping horizon are chosen empirically through the ablation study on the leader tracking task shown in \Cref{tab:tracking-horizon,tab:mapping-horizon}. We fix the mapping horizon to \unit[0.5]{s} for the tracking study and the tracking horizon to \unit[1]{s} for the mapping study.

    \begin{table}
    \centering
    \caption{Ablation study on the tracking horizon in seconds, the full-body error in centimeters and foot contact in F1.}
    \small
    \begin{tabular}{lcc}
        \toprule
        Tracking horizon & Full-body err. $\downarrow$ & Foot contact $\uparrow$ \\
        \midrule
        0.5              & 10.63          & 0.78         \\
        1.0              & 8.32           & 0.90         \\
        1.5              & 11.04          & 0.85         \\
        2.0              & 11.11          & 0.88         \\
        \bottomrule
    \end{tabular}
    \label{tab:tracking-horizon}
\end{table}

    \begin{table}
    \centering
    \caption{Ablation study on the mapping horizon in seconds, the full-body error in centimeters and foot contact in F1.}
    \small
    \begin{tabular}{lcc}
        \toprule
        Mapping horizon & Full-body err. $\downarrow$ & Foot contact $\uparrow$ \\
        \midrule
        0.25            & 14.69          & 0.71         \\
        0.5             & 8.32           & 0.90         \\
        0.75            & 13.25          & 0.79         \\
        1.0             & 13.92          & 0.80         \\
        \bottomrule
    \end{tabular}
    \label{tab:mapping-horizon}
\end{table}

\begin{figure}
    \centering
    \includegraphics[width=\linewidth]{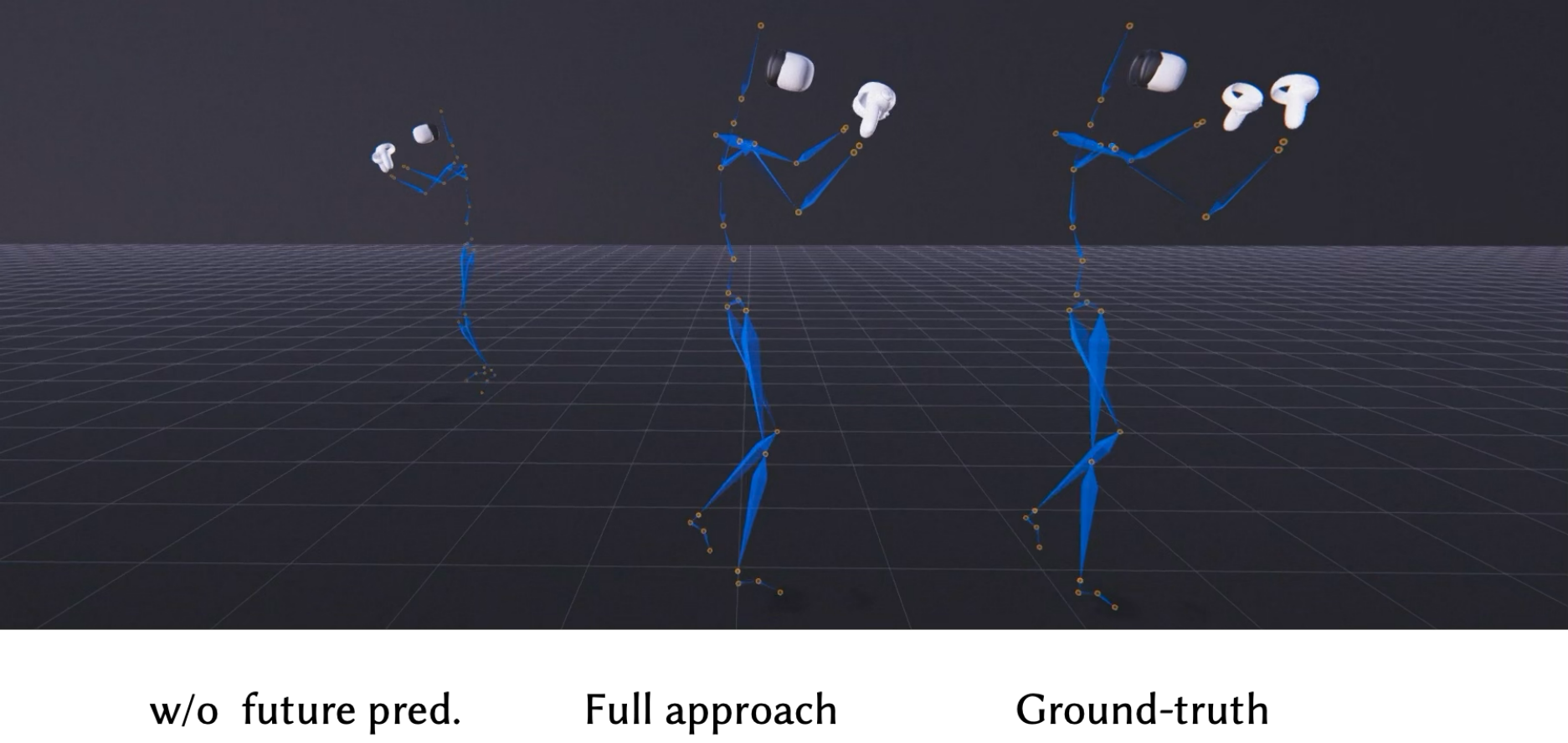}
    \caption{Ablation study on the leader's future prediction. When tracking without the mapping network's future prediction, the autoregressive model is fragile and can quickly diverge.}
    \label{fig:ablation1}
\end{figure}

\paragraph*{Leader's future prediction} We show that for tracking the leader's full-body motion, directly using past three-point trajectories as input leads to less responsive predictions during test motions, resulting in incoherent generated motion, as seen in \Cref{fig:ablation1}. This behavior arises because autoregressive models, when tasked with future prediction, often overfit the training data rather than learning to effectively track input signals. In contrast, when future three-point trajectories are provided, the network becomes more decisive, focusing on extrapolation rather than simply recalling past motions. Note that even though the future three-point trajectories are also predicted by a neural network, the mapping network is not autoregressive, making sure the rollouts are stable.
A qualitative result is included in the accompanying video.

\begin{figure}
    \centering
    \includegraphics[width=\linewidth]{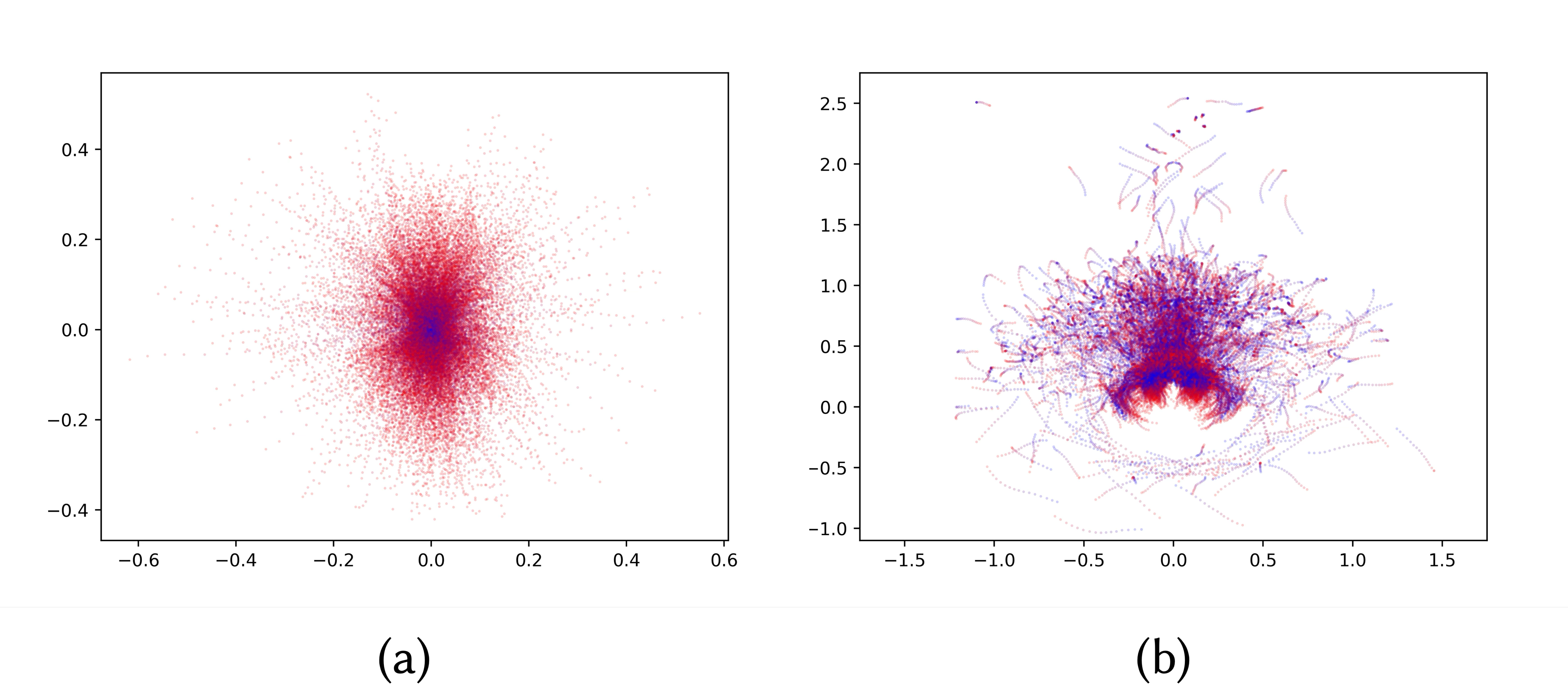}
    \caption{Visualization of future root trajectory input: (a) with the mapping network and (b) without the mapping network. Both are in the coordinate frame of the follower. Each input trajectory starts with blue and ends in red. Note that the scale for (a) and (b) is different. }
    \label{fig:input-sig}
\end{figure}

\begin{figure}
    \centering
    \includegraphics[width=\linewidth]{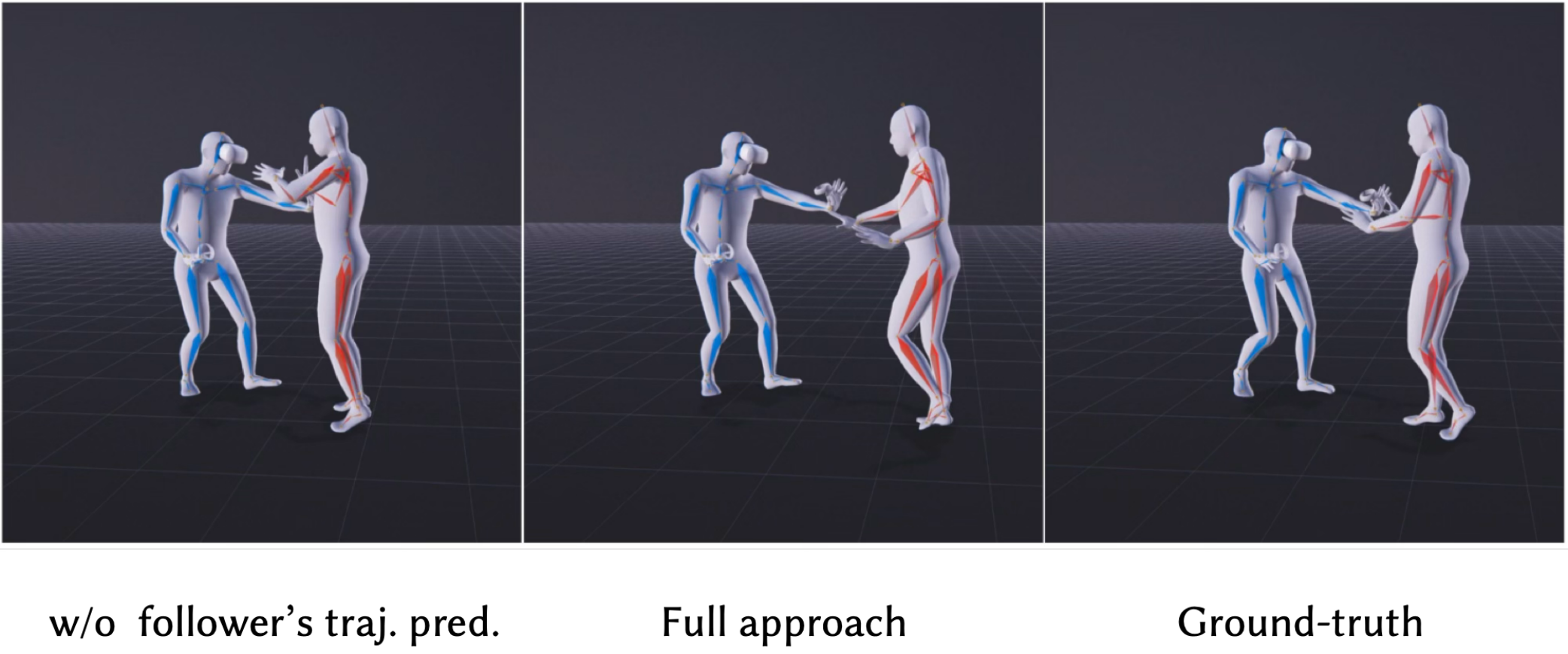}
    \caption{Ablation study on the follower's three-point trajectory prediction. The predicted follower is in red. When directly predicting the follower's full-body motion with the leader's future trajectory, the follower is significantly less responsive due to the input signal distribution.}
    \label{fig:ablation2}
\end{figure}

\paragraph*{Follower's future prediction.} We also show that without the mapping network, it is difficult to synthesize the follower's motion directly from the leader's three-point trajectories. To avoid the aforementioned future prediction problem, we use the future three-point trajectories of the leader as input here. Using leader's three-point trajectories complicates the input signal distribution, as shown in \Cref{fig:input-sig}(b). It consists of many outliers that are difficult to learn with a small dataset. With the mapping network, the input of the future trajectory has a much more even distribution, as shown in \Cref{fig:input-sig}(a). Similarly to the previous case, the coupling of interplay prediction and full-body prediction causes the network to overfit due to the limited amount of data, as can be seen in \Cref{fig:ablation2}.
Please refer to the accompanying video for a detailed result.

\begin{figure}
    \centering
    \includegraphics[width=\linewidth]{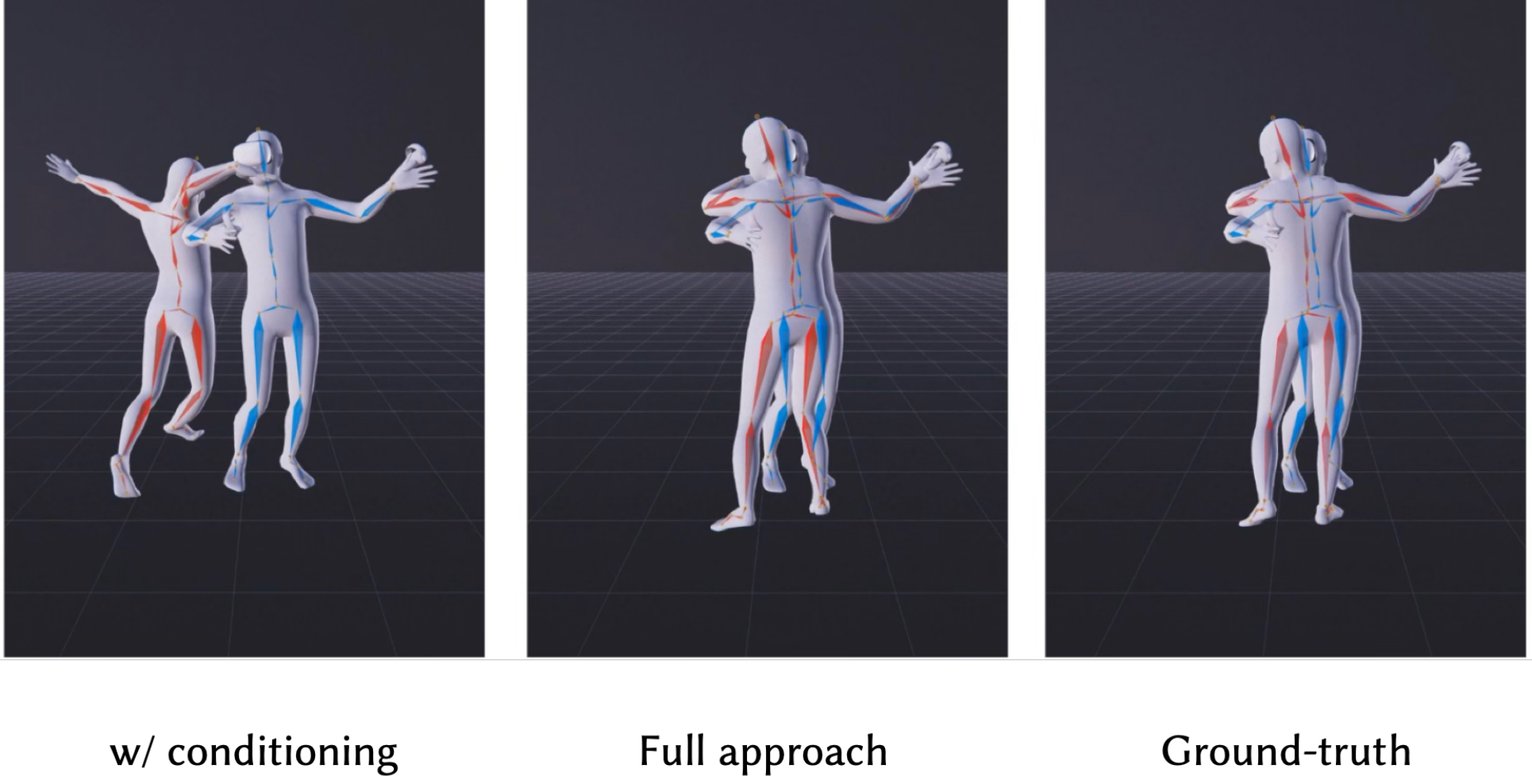}
    \caption{Ablation study of conditioning on the leader's full-body. The predicted follower is in red. When being conditioned, the rollouts become unstable due to the limited amount of training data and the divergence between the predicted leader motion during inference and the ground-truth leader motion used during training.}
    \label{fig:ablation3}
\end{figure}

\paragraph*{Conditioning on leader's full-body.} Although the three-point trajectory is a good descriptor for ballroom dancing motion, it still omits certain details. A natural idea is conditioning the prediction of the follower on the prediction of the leader.
However, this approach faces several challenges: obtaining a reliable future prediction of the leader's motion in real time is difficult, and the additional high-dimensional conditioning can cause the model to overfit. Furthermore, discrepancies between the leader's predicted motion during inference and the ground-truth motion used during training introduce additional difficulty.
It also means we would be conditioning an autoregressive model on the output of another autoregressive model. We show that when leader's full-body motion is used as conditioning, the framework is more fragile and does not generalize well, as seen in \Cref{fig:ablation3} and in the accompanying video.

\begin{table}
    \centering
    \caption{Quantitative evaluations when reversing leader and follower (cm).}
    \small
    \begin{tabular}{lcc}
        \toprule
                 & Leader error $\downarrow$ & Follower error $\downarrow$ \\
        \midrule
        Original & 8.32         & 9.21           \\
        Reversed & 9.05         & 9.79           \\
        \bottomrule
    \end{tabular}
    \label{tab:role-reversal}
\end{table}

\paragraph*{Role reversal.} The leader and the follower roles in ballroom dancing are always clear in practice, and most of the time the leader initiates and the follower maintains the dancing flow. However, we show that from a data-driven perspective, the leader and follower roles are not distinguishable: when we switch the leader and follower roles in the dataset and re-train our model, a similar tracking and full-body prediction error can be achieved, as seen in \Cref{tab:role-reversal}. This may be because the way the dancers communicate (e.g.\ through transmitting forces) is not captured, and the reaction time may be below the target frame time we use in our model.

\section{Discussion}
\label{sec:discussion}

In this work, we present a deep learning framework to synthesize ballroom dancing in various styles from three-point input in real time. Modeling the interplay between two dancers' full body motion is challenging in itself, especially given the risk of overfitting having only a small dataset. The key to our success is exploiting the three-point trajectory as a dancer's motion descriptor.
The low dimensional yet informative representation facilitates the framework structure, allowing us to directly model the interplay on the three-point trajectory level with a multi-layer perceptron, bypassing the dependency of autoregressive models when involving full-body motion dependency, which ensures stable and responsive rollouts. To solve the subsequent three-point tracking problem on ballroom dancing, we adopt a modified version of Codebook Matching with continuous latent space. We additionally demonstrate the effectiveness of the tracking network on the large-scale LaFAN dataset, demonstrating that a deterministic model suffices to produce natural tracking results, though it may diverge from the ground truth in inherently ambiguous scenarios.
With the two networks together, we can synthesize the leader's motion accurately following the three-point input and a plausible follower's motion that synchronizes with the leader.

\subsection{Per-step determinism}

Our framework rests on the observation that, although three-point tracking is globally under-constrained, the per-step prediction can be handled by a deterministic model once the network is conditioned on the current pose and a \emph{future} \tpts trajectory, suggesting the prediction should be close to unimodal. We note that this close-to-unimodal per-step observation is consistent with substantial multi-modality at the rollout level: even a small per-step probability of deviating from the dominant mode compounds multiplicatively over the rollout horizon, so the chance that a trajectory remains on a single mode throughout falls off rapidly with the number of steps, making the distribution over full trajectories far broader than that at any individual transition. The multi-modality of the problem is therefore largely a property of the rollout rather than of any single step. This explanation is empirical, leaving theoretical support and quantification of the multi-modality as an interesting future direction. As the motion data grows more diverse, the per-step distribution becomes more difficult to learn, and a generative model may be required.

\subsection{Limitations}

\begin{figure}
    \centering
    \includegraphics[width=0.8\linewidth]{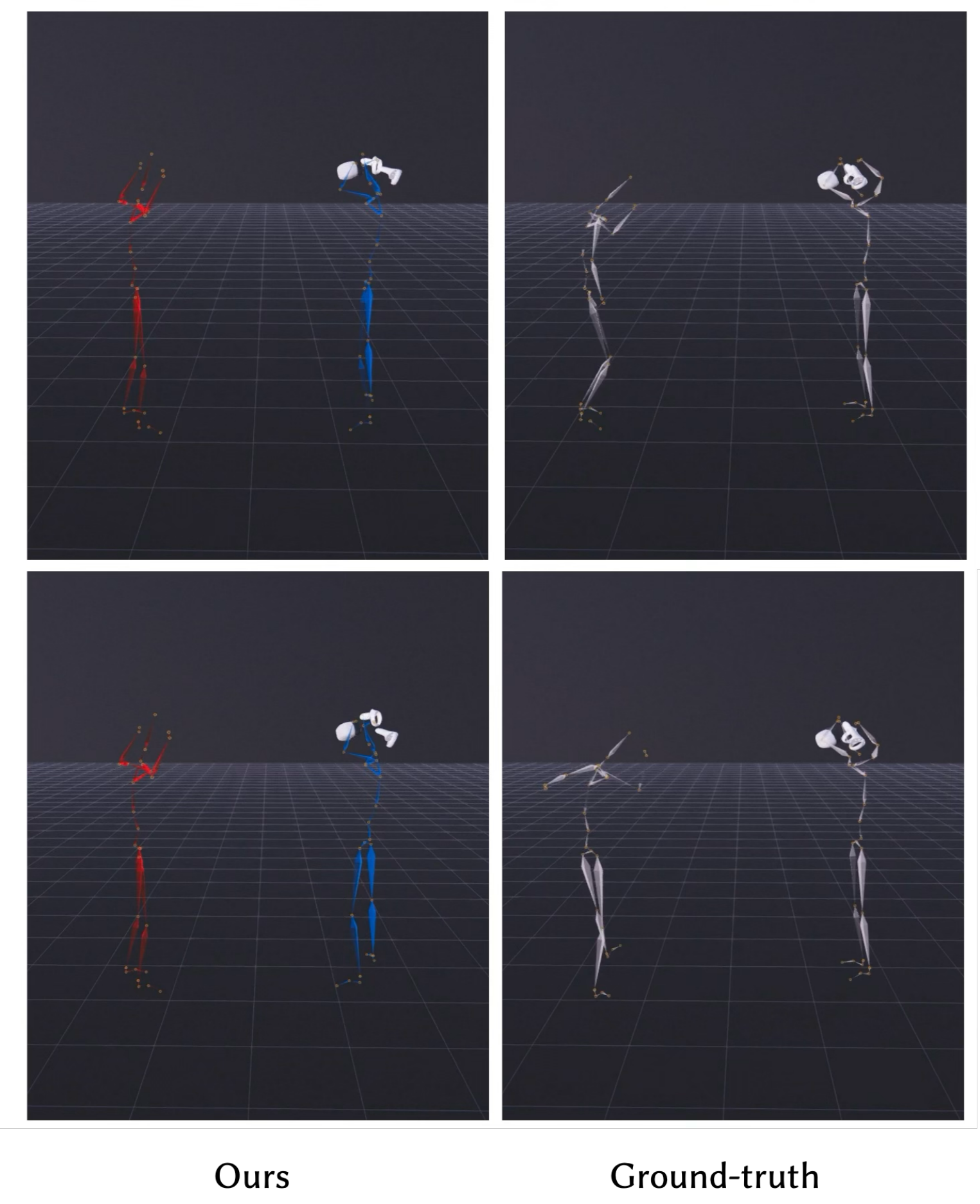}
    \caption{In an example of active following paradigm, the leader remains relatively static while the follower takes initiative. Our model generates static follower motion because of its deterministic nature on a constant input signal.}
    \label{fig:limitation}
\end{figure}

One weak point of our model is the lack of generative ability in the mapping network. This limits our model to handle passive follow only, while failing to model diverse responses in an active follow paradigm, most notably, when the leader stays rather static and the follower responds freely, as illustrated in \Cref{fig:limitation}. An interesting direction would be applying a generative framework such as diffusion to handle such cases.

Related to that, our deterministic tracking formulation is validated empirically on the highly structured dance dataset and the LaFAN dataset, where conditioning on the current pose and the future \tpts trajectory keeps the per-step distribution close to unimodal, so it can be handled by a deterministic model. For substantially more diverse motion repertoires that include sitting, squatting or unconstrained interaction with the environment~\cite{starke2024categorical}, the per-step distribution may become genuinely multimodal, and a generative tracker would likely be necessary.

Our model also relies on head orientation only for calculating root trajectories, which is crucial for making a stable prediction. When the head is facing downwards or upwards close to perpendicular to the floor, the singularity makes the root trajectory calculation unstable and causes unstable tracking results. This issue is primarily revealed in some motion categories in LaFAN, although not presented in dancing datasets.

We handle users of different heights by using the same calibration as codebook matching~\cite{starke2024categorical}, where the model is trained on a single person's data, and the height of the user is used to globally scale the tracker locations to match the height of the training avatar. However, users with significantly different body proportions would require additional data.

Another issue is handling body contact handling. It would be beneficial to adapt techniques like the contact matrix in the work of Starke et al.~\cite{starke2021neural} to enhance the quality of motion. It would also be interesting to explore the possibility of synthesizing both the leader and the follower at the same time. Our current model synthesizes the two dancers independently after jointly predicting the two correlated three-point trajectories. Jointly modeling the two dancers may yield more coherent dancing flows.

\section*{Acknowledgments}
This work was supported in part by the European Research Council (ERC) under the Horizon 2020 Research and Innovation Programme (ERC Consolidator Grant, agreement No.\ 101003104, MYCLOTH). 
Open access publishing facilitated by ETH Zurich, as part of the Wiley-ETH Zurich agreement via the Consortium Of Swiss Academic Libraries.

\bibliographystyle{eg-alpha-doi}
\bibliography{bibs.bib}

\appendix

\section{Network architecture}
\label{appendix:arch}

\begin{table}
    \centering
    \caption{Architecture of the encoder and estimator network.}
    \small
    \begin{tabular}{llc}
        \toprule
        Layers     & Contents                                    \\
        \midrule
        1          & \ttt{White noise} ($\mathcal{N}(0, 0.1)$)   \\
                   & \ttt{Dropout(0.1)}                        & \\
                   & \ttt{Linear}                              & \\
                   & \ttt{ELU}                                 & \\
        \midrule
        2 $\sim$ 5 & \ttt{Dropout(0.1)}                        & \\
                   & \ttt{Linear}                              & \\
                   & \ttt{ELU}                                   \\
        \midrule
        6          & \ttt{Dropout(0.1)}                        & \\
                   & \ttt{Linear}                              & \\
                   & \ttt{Reshape + Softmax}                   & \\
        \bottomrule
    \end{tabular}
    \label{tab:arch-enc}
\end{table}

\begin{table}
    \centering
    \caption{Architecture of the decoder network.}
    \small
    \begin{tabular}{llc}
        \toprule
        Layers     & Contents             \\
        \midrule
        1          & \ttt{Dropout(0.1)} & \\
                   & \ttt{Linear}       & \\
                   & \ttt{ELU}          & \\
        \midrule
        2 $\sim$ 5 & \ttt{Dropout(0.1)} & \\
                   & \ttt{Linear}       & \\
                   & \ttt{ELU}            \\
        \midrule
        6          & \ttt{Dropout(0.1)} & \\
                   & \ttt{Linear}       & \\
        \bottomrule
    \end{tabular}
    \label{tab:arch-dec}
\end{table}

The encoder, decoder, estimator of the tracking network and the mapping networks are designed to be MLPs, with detailed architecture in \Cref{tab:arch-enc} and \Cref{tab:arch-dec}. The input is flattened into a 1D vector before being fed into the network. All the networks use 1024 neurons for the hidden layers.

The vanilla MLP network is constructed by concatenating the encoder and decoder components of the tracking networks and trained with only \Cref{eq:loss-rec}, following the same architecture as described above.

\section{Motion recovery and visualization}
\label{appendix:visualization}

To recover the final motion from the neural network output, we combine the predicted joint positions $\PP$ and joint rotations $\RR$ into per-joint rigid transformations. We use the predicted foot contact labels for foot locking: when a foot is labeled in contact, we solve for the corresponding leg joint rotations by inverse kinematics to fix the foot's global position, which removes residual foot sliding.
We then apply these per-joint transformations to the character geometry using linear blend skinning with the rig's skinning weights to produce the rendered mesh shown in the figures and the accompanying video.
The same recovered motion is used consistently throughout the paper for all qualitative visualizations and quantitative metrics. Using positions alongside rotations bypasses the hierarchical error accumulation of rotation-only forward kinematics and avoids operating purely on the non-linear $\mathrm{SO}(3)$ manifold, while still producing a full rigid transformation per joint that drives skinned rendering directly.

\end{document}